\documentclass[aps,prb,showpacs,amsmath,floatfix,amssymb,preprint]{revtex4}

\usepackage{graphicx}
\usepackage{dcolumn}%
\usepackage{bm}%
\usepackage{color} 

\begin{document}

\title{Strain-induced half-metallic ferromagnetism in zinc blende
CrP/MnP superlattice: First-principles study}

\author{Gul Rahman\footnote[2]{Present Address: Graduate Institute of Ferrous Technology, Pohang
University of Science and Technology, Pohang 790-784, Republic of
Korea.}}
\email{grnphysics@yahoo.com}\

\begin{center}
\textcolor{black}{PHYSICAL REVIEW B \textbf{81}, 134410 (2010)} 
\end{center}

\affiliation{Department of Physics, University of Ulsan,
Ulsan 680-749, Republic of Korea}


\begin{abstract}
Using first-principles calculations within generalized gradient
approximation, the electronic and magnetic properties of
zinc blende (zb) CrP/MnP superlattice are investigated. The
equilibrium lattice constant is calculated to be $5.33\,$\AA. The
stability of ferromagnetic zb CrP/MnP superlattice against
antiferromagnetism is considered and it is found that the
ferromagnetic CrP/MnP superlattice is more stable than the
antiferromagnetic one. It is shown that at the equilibrium lattice
constant the CrP/MnP superlattice does not show any half
metallicity mainly due to the minority $t_{2g}$ states of Cr and
Mn. However, if strain is imposed on the CrP/MnP superlattice then
the minority $t_{2g}$ electrons shift to higher energies and the
proposed superlattice becomes a half-metal ferromagnet. The effect of tetragonal and orthorhombic distortions on the half metallicity of zb CrP/MnP superlattice is also discussed. It is also
shown that InP-CrP/MnP/InP is a true half-metal ferromagnet. 
The half metallicity and magnetization of these superlattices are
robust against tetragonal/ orthorhombic deformation.
\end{abstract}


\maketitle
In the past decade, extensive research has been undertaken to
explore the electron spin degree of freedom for the design of new
electronic devices. This has been motivated by the prospect of
using spin in addition, or as an alternative, to charge as the
physical quantity carrying information, which may change device
functionality to an entirely new paradigm: dubbed spintronics. One
of the key requirements for engineering spintronics devices is the
efficient injection of spin-polarized charge carriers from a
terminal (that is, a ferromagnet electrode) into a semiconductor
interlayer. A half-metal ferromagnet (HMF),\cite{Groot} in which
one spin channel is metallic while the other is semiconducting and
hence 100\%\ spin polarization at the Fermi level ($E_\mathrm{F}$)
is expected, is considered optimal for spintronic devices. After
the discovery of HMF by Groot \textit{ et al.} \cite{Groot} in
1983, much effort has been made to study new half-metallic
compounds. Zinc blende (zb) transition metal pnictides and
chalcogenides in their metastable state 
are among these new half-metallic compounds.~\cite{Ref-gr2,ref-Xu,Ref-gr4,Ref-gr5,Ref-gr6,Ref-gr7,Miao} Much attention was
diverted to these zb compounds after the pioneering work of
Akinaga \textit{ et al.} \cite{Akinaga} who predicted half-
metallic behavior of zb CrAs grown as a thin film. The zb Cr(Mn)
pnictides have been found to hold a large magnetic moment, i.e.,
$3.0$ ($4.0$) $\mu_{B}$ for Cr (Mn) compounds, per formula
unit.\cite{Ref-gul9} The calculated, using the full potential
linearized augmented plane wave (FLAPW) method, equilibrium
lattice constant of bulk zb CrP is found to be $5.35$ \AA \;and
becomes a HMF for lattice constants $\geq 5.48$ \AA~\cite{Ref-gr7} ($a_\textrm{CP}$).
On the other hand the equilibrium lattice constant of zb MnP was
calculated to be $5.308\,$\AA, using the FLAPW, and was shown that
zb MnP is not a HM at the equilibrium lattice
constant.\cite{Ref-gul9} Note that, in nature, these zb type materials are stable in structures other than zb which do not  show any half-metallic behavior.~\cite{Ref-gul9,ref-Xu,Sanvito,Hong,Ravindran} The previous theoretical calculations showed that the energy difference between zb and the natural structure, such as MnP and NiAs, are about $0.2$--$1.0$ {eV}. However, it is argued that under eptixal growth condition, most of the Cr and Mn pnictides and chalcogenides, the natural structure is either always lower in energy than zb or only becomes higher in energy under a lattice expansion that is larger than the ones that can be achieved by epitaxy on suitable semiconductor substrates.\cite{alex} The ground state structure of MnP and CrP is  orthorhombic (MnP-type)\cite{Ref-gul9} and attention is paid explicitly to zb type superlattices and to search for half metallicity in this superlattice.

In many cases, the spin polarization at $E_\mathrm{F}$ is
sensitive to structural deformation and consequently the proposed
HM often loses the high spin polarization in atomic disordered
states at surface, interface, and other structures.\cite{wijs}
The high degree of spin polarization at the surface and interface
is crucial for realistic applications. More recently, we
calculated the magnetism of zb CrP (001) surface for different
lattice constants and no surface states were found for the
Cr-terminated surfaces, while the P-terminated surfaces destroyed
the half-metallic behavior of zb CrP.\cite{Rahman} The purpose of
the present calculations is to search for a new HM layered
material based on zb structure, i.e., CrP/MnP, and to see the effect of strain on electronic and magnetic properties. Here, it is shown that zb CrP/MnP
superlattice does not show any half metallicity at its equilibrium
lattice constant, but shows half-metallic behavior when external
strain is applied.

\begin{figure}
\begin{center}
\includegraphics[width=\columnwidth]{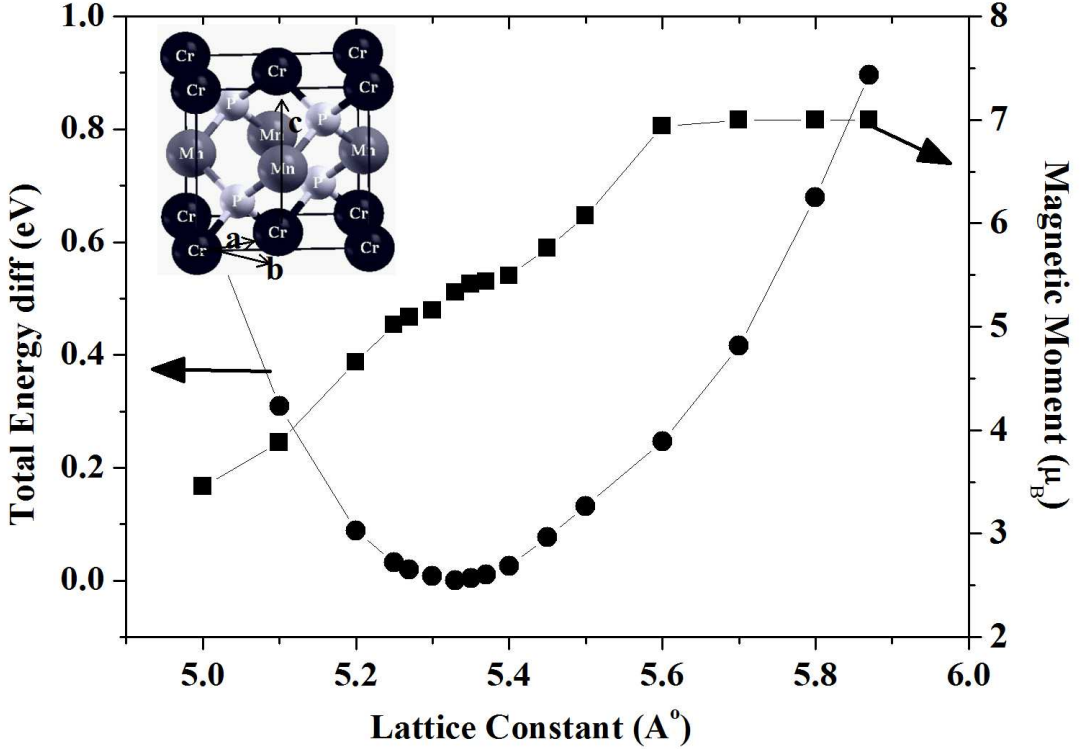}
\caption{ The calculated total energy (eV) and magnetic moment
($\mu_\mathrm{B}$) vs lattice constant (\AA) of zb CrP/MnP
superlattice. The filled circles (left panel) and squares (right
panel) show the energies and magnetic moments. The inset on the
top shows the conventional unit cell of the zb CrP/MnP
superlattice and the arrows with labels {$a,b,$ and $c$} represent
the tetragonal unit cell. The total energy is given with respect to the equilibrium structure.} \label{TEfig}
\end{center}
\end{figure}

The spin-polarized FLAPW method \cite{flapw} based on density
functional theory in the generalized gradient approximation
(GGA)\cite{GGA} was used for first-principles calculations. The
core electrons were treated fully relativistically, whereas the
valence electrons were treated semirelativistically. The basis
functions were expanded in terms of spherical harmonics up to
$l\le8$ inside each muffin-tin sphere and plane waves in the
interstitial region, respectively. The muffin-tin radii were taken
to be $2.20$ a.u. for each atom in the calculations. Convergence
with respect to the basis functions and $k$ points was carefully
checked. The layered structure of zb CrP/MnP considered in these
calculations is shown in the inset of Fig.~\ref{TEfig}. The unit
cell is tetragonal with lattice constant $c=\sqrt{a}$ and is
highlighted with arrows. In the tetragonal primitive cell there
are one Cr, one Mn, and two P atoms.


The equilibrium lattice constant of zb CrP/MnP layered structure
is determined from the total energy calculations. To search for
the equilibrium  lattice constant, this parameter is varied from
$\sim 5.0$\AA \;to $\sim 6.0$\AA. A global energy minimum in the
ferromagnetic (FM) state (see Fig.~\ref{TEfig}) with lattice
constant of $5.33\,$\AA ($a_\textrm{eq}$), which is close to the average
of the equilibrium lattice constants of zb CrP and MnP, was found.
To consider the stability of FM against the antiferromagnetic(AFM)
configuration, a conventional unit cell (eight atoms unit cell)
was used. It is also possible to use a tetragonal unit cell to
consider only one type of AFM structure. However, two different
types of AFM ordering, i.e., AFM coupling between Cr or Mn atoms
(denoted as AFM1) and AFM coupling between Cr and Mn atoms
(denoted as AFM2), were considered and a conventional unit cell
was used. The FM state was found to be more stable than the AFM1
(AFM2) state by $0.71$ ($0.46$) {eV}, which ensures that the true
ground state of CrP/MnP superlattice is FM. The electronic
structures were  calculated for all lattice constants and very
interestingly it was found that the CrP/MnP layered structure is a
HMF for lattice constants $\ge 5.60\,$\AA ($a_\textrm{HM}$), which will
be discussed below.
\begin{figure}
\begin{center}
\includegraphics[width=\columnwidth]{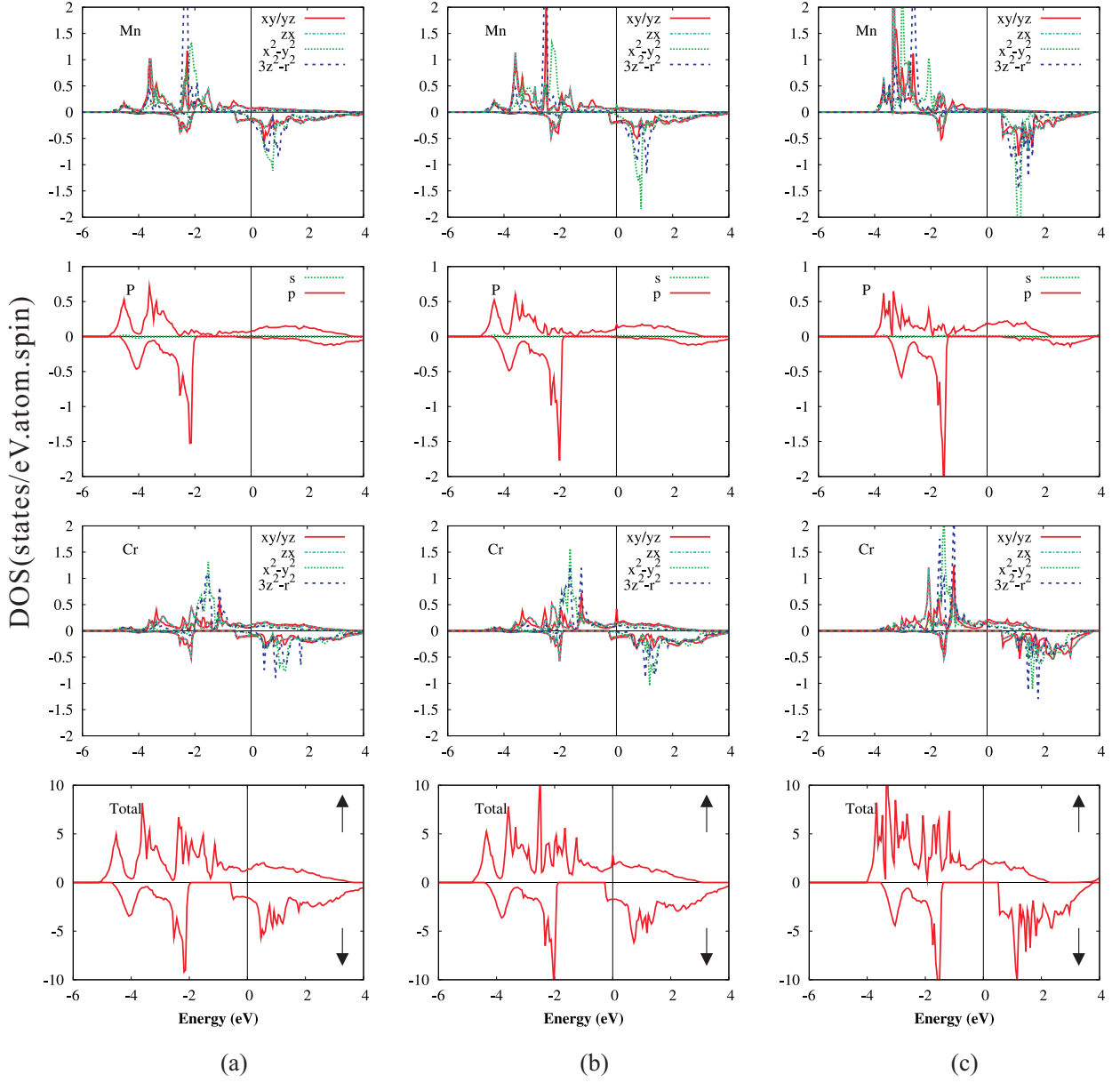}
\caption{ The calculated total and atom projected densities of
states of the zb CrP/MnP superlattice at (a) $a_\textrm{eq}$, (b) $a_\textrm{CP}$ and (c)
$a_\textrm{HM}$. The solid, dasehd-dotted, dotted, and dashed lines show the $xy/yz, zx, x^2-y^2$, and $3z^2-r^2$ states, respectively. The thin solid and dotted lines
represent the $s$ and $p$ states of P. The solid lines at the bottom panel
show the total density of states. The up (down) arrows show the
majority (minority) spins, and the Fermi level ($E_\textrm{F}$) is set to
zero.} \label{dosfig}
\end{center}
\end{figure}

The calculated total and atom projected density of states (DOS) at
$a_{eq}$ are shown in Fig.~\ref{dosfig}(a). The Cr (Mn) $d$ bands
are decomposed into $t_{2g}$($xy,yz,zx$) and $e_{g}$ ($x^2-y^2,3z^2-r^2$) states. The bands at
$\sim -10.0\,${eV} (not shown) have anion $s$ character and are
well isolated from the bonding states so that they do not
participate in bonding. The total DOS shows that the CrP/MnP
superlattice is not a half-metal due to a finite DOS at
$E_\mathrm{F}$ in the minority-spin states. The majority spin
states in the interval $-5.0$ to $-2.5$ eV are bonding states and
are formed by the Cr- and Mn- $t_{2g}$ and P $p$ electrons. The
$d$ nonbonding states, which are mainly formed by the $e_{g}$
electrons of Cr and Mn, are located between $\sim -2.5\,${eV} and
$-1.2$ {eV}. The antibonding states are around $-1.0\,${eV} and
extend toward $E_\mathrm{F}$. The minority spins have hybrid
bands between $-5.0\,${eV} and $-2.0\,${eV}, and the strong
bonding peak $\sim -2.0\,${eV} is contributed by the $t_{2g}$
electrons of Mn and Cr and the $p$ electrons of P. The antibonding
minority spins are located below $E_\mathrm{F}$. The DOS at
$E_\mathrm{F}$ is very large due to the minority spins which makes
the CrP/MnP superlattice a metal. One can easily see that the
CrP/MnP DOS is a mixture of Cr/Mn 3$d$ states and the Mn states,
which dominate the $d$ manifold at lower energies. The Cr states
dominate at higher energies, which is consistent with the nuclear
charges. The $t_{2g}$ states of Cr and Mn are responsible for the
absence of half metallicity in the CrP/MnP superlattice. This
happens because the exchange splitting between the
bonding-antibonding is not strong at $a_\textrm{eq}$ to push the minority
spins to higher energies, which can open a band gap in the
minority spin states. The DOS of zb CrP/MnP superlattice was also calculated at $a_\textrm{CP}$ [see Fig.~\ref{dosfig} (b)], which is larger than $a_\textrm{eq}$ . It is noticeable that all the features of DOS of zb CrP/MnP superlattice at $a_{CP}$ are similar to Fig.~\ref{dosfig}(a), i.e., absence of half metallicity. The DOS shows that with increasing the lattice constant, the minority conduction band moves to higher energy and $E_\mathrm{F}$ lies just above the conduction band minimum. This indicates the possibility of half metallicity in zb CrP/MnP superlattice.

The absence of half metallicity can be healed by volume expansion
or by applying strain, which is not only used to control the
carrier density, but also the mobility of the carriers.\cite{GdN-ref}
Strain is also considered to be a source to modify
the electronic structure, magnetic and structural properties of
materials.\cite{Falicov} As the lattice volume was increased
(applying volumetric strain), it was found that the CrP/MnP superlattice
developed insulating features in the minority spin states and the
magnetic moment was also increased.

The calculated atom projected and the total DOS at $a_\textrm{HM}$ are
shown in Fig.~\ref{dosfig}(c). The DOS clearly shows that the
CrP/MnP superlattice is a HM and now there are no minority states
at $E_\mathrm{F}$. For the majority band, the hybridization
between the Mn $t_{2g}$ and P $p$ dominates between $\sim -4.0$
and $-2.0\,${eV} whereas above $-2.0\,${eV}, the Cr $t_{2g}$ and P
$p$ bands are strongly hybridized. For the majority spins the
conduction in the majority states is mainly dominated by the Cr
$d$ states. The minority spins exhibit a similar structure and the
strong narrow bonding peak at $\sim -1.80$ eV is mainly
contributed by the Cr $t_{2g}$ and P $p$. The minority $d$ states
are shifted significantly relative to the majority $d$ states by
the exchange splitting and the $d$ bands move to antibonding
states, which open a gap ($\geq 1.0$ eV) that is important for
spintronics. The states, which were formed between $\sim-5.0$ and
$-4.0$ at $a_\textrm{eq}$, are shifted to higher energies due to strain.
It is to be noted the $t_{2g}$ states are completely occupied and
the $e_{g}$ states are empty. Comparing the effect of strain on
CrP/MnP superlattice [Fig.~\ref{dosfig}(a) and (b)], one can see
that the Cr/Mn $3d$ majority bands are narrowed by $\sim1.0\,${eV}
and the exchange splitting is increased for the Mn $d$ bands. This
large exchange splitting is accompanied by a large magnetic moment
of the CrP/MnP superlattice (Fig.~\ref{TEfig}). Nevertheless, atomic disorder/antisite defects may affect the half metallicity.\cite{Galakins}

\begin{figure}
\begin{center}
\includegraphics[width=\columnwidth]{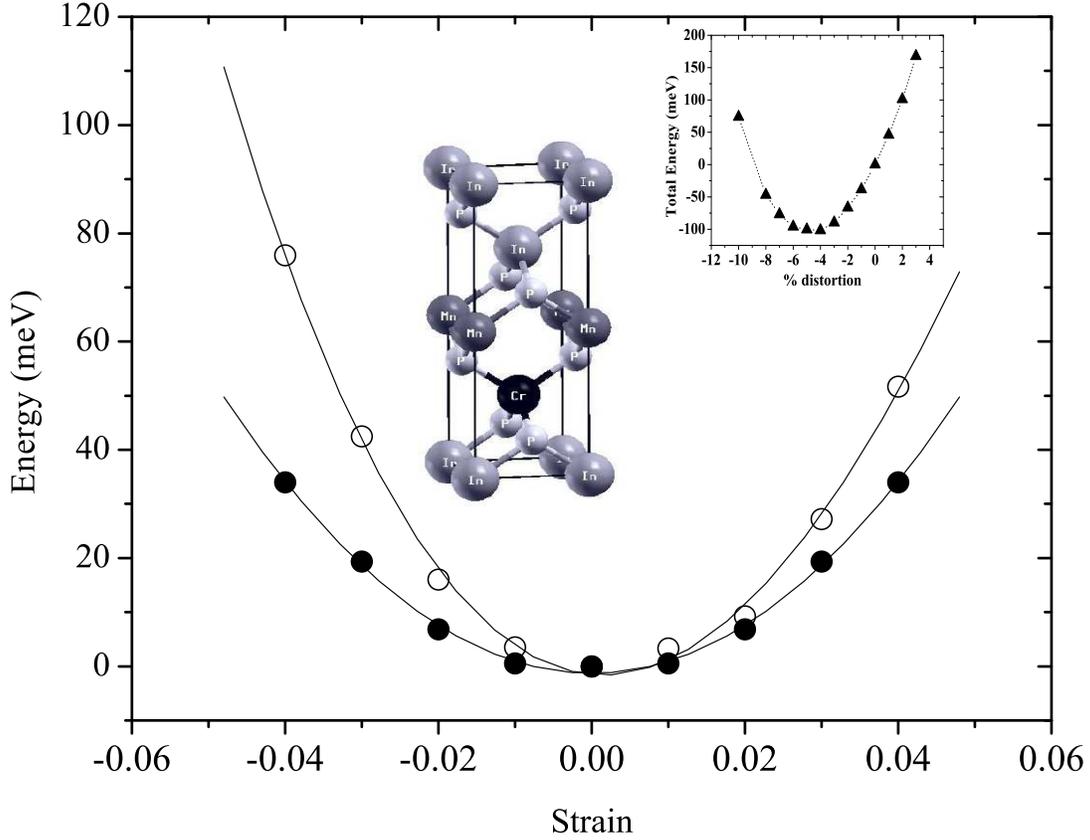}
\caption{ Energy (meV) versus volume conserved tetragonal (open circles) and orthorhombic (filled circles) strains of zb CrP/MnP superlattice.  The inset on the right top shows the total energy (meV) vs tetragonal distortion (\%). The structure of InP-CrP/MnP/InP is also shown in the inset. The total energy is shown with respect to undistorted superlattices. } \label{tetrago}
\end{center}
\end{figure}

To further emphasis on the half-metallic behavior of zb CrP/MnP at $a_\textrm{HM}$, calculations were carried out on the effect of tetragonal and orthorhombic deformations on half metallicity. Volume conserved (Poisson's ratio$=0.5$) distortions ($\pm4\%$) were applied (see Fig.~\ref{tetrago}). It is clear to see that the half metallicity of zb CrP/MnP superlattice is robust to such distortions and the magnetic moment per cell remained constant. Similar behavior was also observed in other metastable zb CrAs and CrSb materials\cite{Shi} which have been grown successfully.\cite{Akinaga,Ohno}

From the total energy and DOS calculations, it is concluded that
it is necessary to apply strain for a transition FM metal to
become a FM half-metal. The proposed strain may be produced if the
CrP/MnP superlattice is grown on a substrate with lattice constant
$> a_\textrm{eq}$. The strain will be generated by the lattice mismatch
between $a_\textrm{eq}$ and the substrate that will weaken the
hybridization between the atoms. An example of material that can
be used for this purpose is InP, whose experimental lattice
constant of $5.87\,$\AA ($a_{\rm{InP}}$) is larger than $a_\textrm{eq}$
of the CrP/MnP superlattice. If CrP/MnP is grown on the InP
substrate, which will induce some hydrostatic strain, then it is
expected that the CrP/MnP will become a HMF. 

\begin{figure}
\begin{center}
\includegraphics[width=0.8\columnwidth]{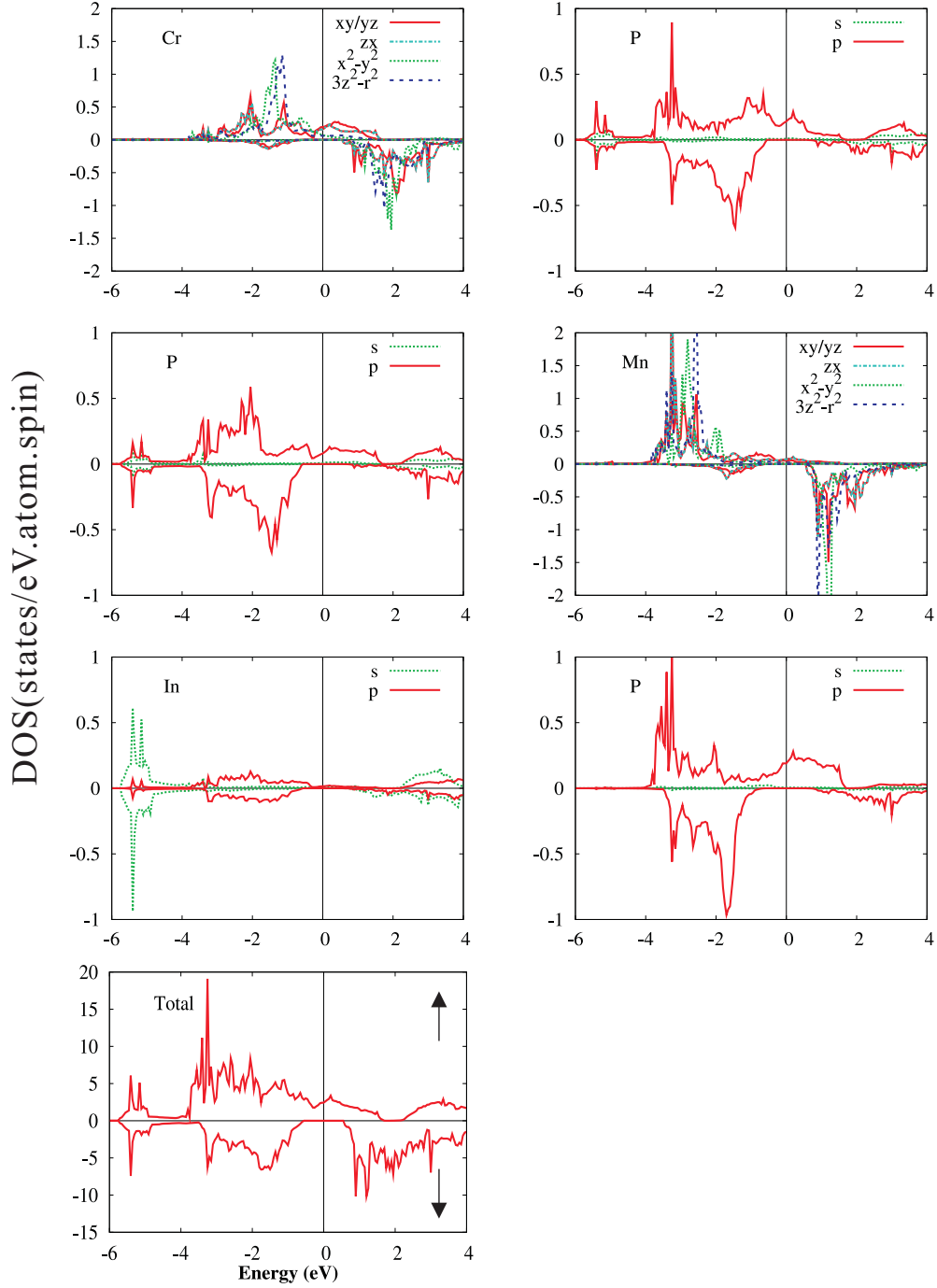}
\caption{The calculated total and atom projected density of states
of InP-CrP/MnP/InP at $a_{\rm{InP}}$. The solid, dasehd-dotted, dotted, and dashed lines show the $xy/yz, zx, x^2-y^2$, and $3z^2-r^2$ states, respectively. The thin solid and dotted lines represent the $s$ and $p$ states of P.
The solid lines at the bottom panel show the total density of
states. The up (down) arrow shows majority (minority) spins and the
Fermi level ($E_\textrm{F}$) is set to zero.} \label{dostrilayer}
\end{center}
\end{figure}

It would be interesting to investigate the half metallicity of multilayer system, i.e., InP/MnP/CrP/InP, which can also be denoted as In$_{0.5}$Cr$_{0.5}$P/ MnP /InP or InP-CrP/ MnP/ InP following the notation of Qian \textit{et al.}~\cite{qian} Further calculations were carried out for InP-CrP/MnP/InP
superlattice (inset of Fig.~\ref{tetrago}) and it was found that
this superlattice is also a HMF. The DOS of InP-CrP/MnP/InP system
is shown in Fig.~\ref{dostrilayer}. The main features (half
metallicity) are identical to the DOS at $a_\textrm{HM}$ except a small
band between $-6.0$ and $-4.0$ eV, which is mainly contributed by
In atoms. More importantly, this  system also has a band gap
($\sim 1.0\,${eV}) in the minority spin states, but it is
decreased compared with Fig.~\ref{dosfig}(c).

It is shown that the CrP/MnP superlattice can be a HMF even if it forms
multilayer structure. Further, the effect of volume conserved tetragonal and orthorhombic distortions on half metallicity was also considered. It was found that the half metallicity is not affected by such distortions, similar to zb CrP/MnP superlattice. Then  the stability of the half-
metallic state of InP-CrP/MnP/InP against tetragonal
distortion(tensile strain), which can be expected in real
applications, is examined. A tetragonal distortion $\pm\,$ \% in
the $c$ direction was applied while keeping the in-plane lattice
constant to find the energy minimum and see the effect of tensile
strain on the magnetic moment and half metallicity. The calculated
energy curve is shown in the inset of Fig.~\ref{tetrago}. The
tetragonally distorted InP-CrP/MnP/InP has an energy minimum at
about 4.5\% compression. The corresponding energy change is $\sim
0.10\,${eV} with respect to the undistorted case. There are some reports where lattice mismatched($\sim 5\%$ ) superlattices are achieved successfully.\cite{Gunshor,Fuj, Will,Schmid}  More
interestingly, the magnetic moment and the half-metallic behavior
of InP-CrP/MnP/InP was not much affected by compression and
elongation. When a zb half-metallic compound is grown on a
semiconductor substrate [on (001)], its lattice will be
tetragonally distorted so that it can assume the in-plane lattice
constant of the semiconductor while approximately keeping its own
atomic volume. Therefore, during the tensile strain the in-plane
bond length between Cr or Mn and P atoms is the same as that of
$a_{\rm{InP}}$. The magnetic moments are already saturated at
$a_{\rm{InP}}$ and the materials show half-metallic property
(Fig.~\ref{TEfig}). The magnetic moment remains constant
($7.0\mu_{B}$) and the DOS for each case shows that the minority
spin has a non-zero gap. The tetragonal distortion can have a
strong effect on the half-metallicity if $E_\mathrm{F}$ lies at
the edge of the gap. For InP-CrP/MnP/InP, $E_\mathrm{F}$ does
not lie at the edge and then the half metallicity is not much
affected. It is to be noted that if the distortion is too large,
the half metallicity can be destroyed. These findings are
encouraging because the interaction of CrP/MnP superlattice with InP
will not cause loss of the half-metallic behavior. Furthermore, no
interface states between CrP/MnP and InP were observed.

Finally, the total magnetic moment per unit cell for the CrP/MnP
is also calculated and it is observed that the magnetic moment
increases with the lattice constant, as shown in Fig.~\ref{TEfig}.
The increase in the magnetic moment with the lattice constant is
due to the increase in bond length (weak hybridization coming from
a reduced $p$-$d$ coupling) and is also observed in other zb
materials.\cite{Sanvito} The magnetic moment becomes saturated
beyond $5.60\,$\AA\;and presents an integer value of $7.0\mu_{B}$,
which is the sum of the free magnetic moments of
Cr($3.0\,\mu_{B}$) and Mn($4.0\,\mu_{B}$).

The proposed material is a HMF at $a_\textrm{HM}$ and shares a common
minority electronic structure with the constituent compounds,
i.e., zb CrP and zb MnP, where $p$-$t_{2g}$ hybrid orbital are
filled and $e_{g}$ states are empty. The magnetic moment per
formula unit ($M$) of the half-metallic {zb} transition metal
pnictides superlattice is given by\cite{Fong}

\begin{align}
M=(Z_{t}-8N_{\rm{P}}) \label{equat}
\end{align}

\noindent where $Z_{t}$ is the total number of valence electrons
in the unit cell and $N_{\rm{P}}$ is the number of P atoms per
unit cell. For the CrP/MnP superlattice, $Z_{t}\,= 23$ and
$N_{\rm{P}}\,=2$ and hence Eq.\ref{equat}  predicts a total magnetic
moment of $7.0\,\mu_{B}$ per unit cell, which is confirmed by
first-principles all-electron FLAPW calculations.

In summary, first-principles calculations, using the FLAPW method
within the GGA for the exchange and correlation potential, were
performed to investigate the magnetic and electronic structures of
zb CrP/MnP superlattices. The equilibrium lattice constant
($5.33\,$\AA) was determined from the total energy calculations.
The stability of ferromagnetism against antiferromagnetism was
also considered and it was shown that the ferromagnetic zb CrP/MnP
superlattice is more stable than the antiferromagnetic
superlattice. The electronic structures revealed that the zb
CrP/MnP superlattice is not a true half-metal and the absence of
half metallicity was discussed in terms of exchange splitting of
the $d$ states of Cr and Mn. However, the zb CrP/MnP superlattice
became half-metal when strain was imposed. The half metallicity of strained CrP/MnP superlattice was not destroyed by tetragonal and orthorhombic deformations. The magnetization and
half-metallic behavior of zb InP-CrP/MnP/InP was also discussed
and it was found that the magnetic and electronic properties of
this superlattice are not much affected by tetragonal distortion.

The author thanks Soon Cheol Hong  and In Gee Kim for useful discussions. This work was partially supported by the Steel Innovation Program by POSCO through POSTECH and the Basic Science Research Program (Grant No. 2009-0088216) through the National Research Foundation funded by Ministry of Education, Science and Technology of Republic of Korea.

\end{document}